\documentclass[aps, prl, amsmath,amssymb,amsfonts,twocolumn,superscriptaddress,floatfix,footinbib,altaffillsymbol]{revtex4-1}

\pdfoutput=1
\usepackage{amsmath}
\usepackage{amssymb}
\usepackage{dcolumn}% Align table columns on decimal point
\usepackage[final]{graphicx}
\usepackage{upgreek}
\usepackage{color}
\usepackage[english]{babel}

\begin{document}
	
\title{Polariton superfluids reveal quantum hydrodynamic solitons}
\author{A.~Amo}
\affiliation{Laboratoire Kastler Brossel, Universit\'e Pierre et Marie Curie, \'Ecole Normale Sup\'erieure et CNRS, UPMC case 74, 4 place Jussieu, 75005 Paris, France}
\affiliation{CNRS-Laboratoire de Photonique et Nanostructures, Route de Nozay, 91460 Marcoussis, France}

\author{S.~Pigeon}
\affiliation{Laboratoire Mat\'eriaux et Ph\'enom\`enes Quantiques, UMR 7162, Universit\'e Paris Diderot-Paris 7 et CNRS, 75013 Paris, France}

\author{D.~Sanvitto}
\affiliation{NNL, Istituto Nanoscienze - CNR, Via Arnesano, 73100 Lecce, Italy}

\author{V.~G.~Sala}
\affiliation{Laboratoire Kastler Brossel, Universit\'e Pierre et Marie Curie, \'Ecole Normale Sup\'erieure et CNRS, UPMC case 74, 4 place Jussieu, 75005 Paris, France}

\author{R.~Hivet}
\affiliation{Laboratoire Kastler Brossel, Universit\'e Pierre et Marie Curie, \'Ecole Normale Sup\'erieure et CNRS, UPMC case 74, 4 place Jussieu, 75005 Paris, France}

\author{I.~Carusotto}
\affiliation{INO-CNR BEC Center and Dipartimento di Fisica, Università di Trento, via Sommarive 14, I-38123 Povo, Italy}

\author{F.~Pisanello}
\affiliation{Laboratoire Kastler Brossel, Universit\'e Pierre et Marie Curie, \'Ecole Normale Sup\'erieure et CNRS, UPMC case 74, 4 place Jussieu, 75005 Paris, France}
\affiliation{NNL, Istituto Nanoscienze - CNR, Via Arnesano, 73100 Lecce, Italy}
\affiliation{Scuola Superiore ISUFI, Università del Salento, Via Arnesano, 73100 Lecce, Italy}

\author{G.~Lem\'enager}
\affiliation{Laboratoire Kastler Brossel, Universit\'e Pierre et Marie Curie, \'Ecole Normale Sup\'erieure et CNRS, UPMC case 74, 4 place Jussieu, 75005 Paris, France}

\author{R.~Houdr\'e}
\affiliation{Institut de Physique de la Mati\`ere Condens\'ee, Facult\'e des Sciences de Base, b\^atiment de Physique, Station 3, EPFL, CH-1015 Lausanne, Switzerland}

\author{E.~Giacobino}
\affiliation{Laboratoire Kastler Brossel, Universit\'e Pierre et Marie Curie, \'Ecole Normale Sup\'erieure et CNRS, UPMC case 74, 4 place Jussieu, 75005 Paris, France}

\author{C.~Ciuti}
\affiliation{Laboratoire Mat\'eriaux et Ph\'enom\`enes Quantiques, UMR 7162, Universit\'e Paris Diderot-Paris 7 et CNRS, 75013 Paris, France}

\author{A.~Bramati}
\affiliation{Laboratoire Kastler Brossel, Universit\'e Pierre et Marie Curie, \'Ecole Normale Sup\'erieure et CNRS, UPMC case 74, 4 place Jussieu, 75005 Paris, France}

\begin{abstract}
\bf{A quantum fluid passing an obstacle behaves differently from a classical one. When the flow is slow enough, the quantum gas enters a superfluid regime and neither whirlpools nor waves form around the obstacle. For higher flow velocities, it has been predicted that the perturbation induced by the defect gives rise to the turbulent emission of quantised vortices and to the nucleation of solitons. Using an interacting Bose gas of exciton-polaritons in a semiconductor microcavity, we report the transition from superfluidity to the hydrodynamic formation of oblique dark solitons and vortex streets in the wake of a potential barrier. The direct observation of these topological excitations provides key information on the mechanisms of superflow and shows the potential of polariton condensates for quantum turbulence studies.}
\end{abstract}

\maketitle

Superfluidity is the remarkable property of flow without friction (1). It is characterised by the absence of excitations when the fluid hits a localised static obstacle at flow speeds $v_{flow}$ below some critical velocity $v_{c}$. For small potential barriers, the critical velocity is given by the Landau criterion as the minimum of $\omega(k)/k$ , with $\omega(k)$ being the dispersion of elementary excitations in the fluid. In the case of dilute Bose-Einstein condensates (BECs), $v_{c}$ corresponds to $c_{s}$, the speed of sound of the quantum gas. For supersonic flows ($v_{flow}>c_{s}$), small obstacles induce dissipation (drag) via the emission of sound waves (2, 3).

When the barrier is big, larger than the fluid’s healing length –the minimum distance induced by particle interactions for changes in the density of the condensate, the density modulations caused by the barrier can generate topological excitations, such as vortices and solitons. These quantum hydrodynamic effects have been predicted to reduce the critical velocity (4, 5). 

Despite the amount of theoretical work (4-6), a limited number of experimental studies have addressed hydrodynamic features in atomic condensates through the observation of the break up of superfluidity at fluid velocities lower than the speed of sound (7, 8). Solitons in a quasi-one dimensional geometry (9) and the nucleation of vortex pairs in an oblate BEC have been reported (10, 11). Far from the hydrodynamic regime, formation of vortices and solitons has been shown by engineering the density and phase profile of the atomic condensate (12, 13), or by the collision of two condensates (14).

Polariton superfluids appear promising in view of quantitative studies of quantum hydrodynamics. Polaritons are two-dimensional composite bosons arising from the strong coupling between quantum well excitons and photons confined in a monolithic semiconductor microcavity. They possess an extremely small mass $m_{pol}$ on the order of $10^{-8}$ that of hydrogen, which allows for their Bose-Einstein condensation at temperatures ranging from few kelvins (15) up to room temperature (16). All parameters of the system such as the flow velocity, density, and shape and strength of the potential barriers can be finely tuned with the use of just one (3) or two (17) resonant lasers, and by sample (18) or light induced engineering (19). A crucial advantage with respect to atomic condensates is the possibility of fully reconstructing both the density and the phase pattern of the polariton condensate from the properties of the emitted light (20). This has been exploited in the recent observations of macroscopic coherence and long range order (15, 18, 21), quantised vortices (20), superfluid flow past and obstacle (3, 17, 22) and persistent superfluid currents (23). 

Here we use a polariton condensate to reveal quantum hydrodynamic features, whereby dark solitons and vortices are generated in the wake of a potential barrier. Following a recent theoretical proposal (24), we investigate different regimes at different flow speeds and densities, ranging from superfluidity to the turbulent emission of trains of vortices, and the formation of pairs of oblique dark solitons of high stability. For spatially large enough barriers, soliton quadruplets are also observed.

\begin{figure*}[t]
	{\includegraphics[width=0.7 \textwidth]{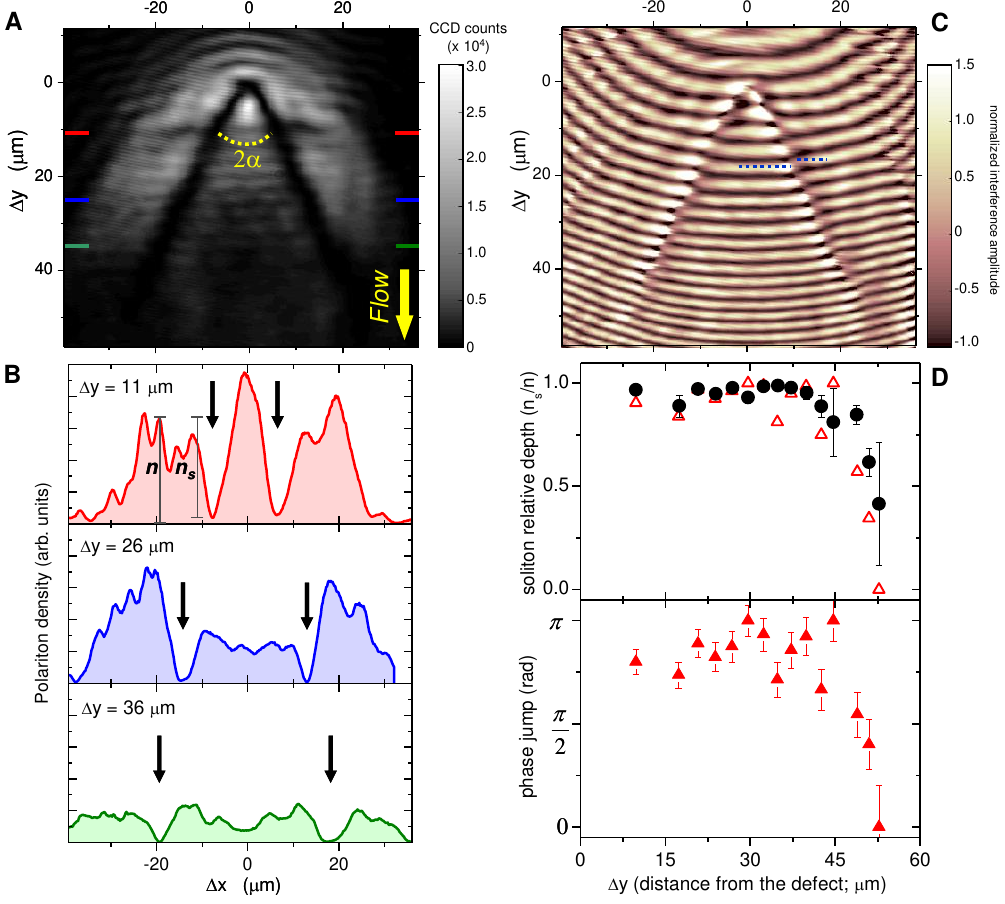}}
	\caption{
(A) Real space emission showing a soliton doublet nucleated in the wake of a photonic defect located at the origin. (B) Horizontal profiles at different downflow distances from the defect $\Delta y$. Arrows indicate the soliton position. (C) Interference between the emitted intensity and a constant phase reference beam, showing phase jumps along the solitons (dashed lines). The curved shaped of the fringes and the decreasing interfringe distance arise from the geometry of the reference beam. (D) Soliton depth (black dots) and phase jump obtained from (C) (full triangles; see Fig. 8), showing a strong correlation. Open triangles: soliton depth obtained from the measured phase jump and Eq. 1.
}
	\label{Fig1}
\end{figure*}

Our experiments are performed in an InGaAs/GaAs/AlGaAs microcavity at 10~K (25). We excite the system with a continuous wave single mode laser quasi-resonant with the lower polariton branch at an angle of incidence $\theta$, resulting in the injection of a polariton fluid with a well defined in plane wavevector (3) ($k=k_0 \sin(\theta)$, where $k_0$ is the wavevector of the excitation laser field) and velocity $v_{flow}=k\hbar/m_{pol}$. The speed of sound of the fluid $c_{s}$ is related to the polariton density $\left|\psi\right|^2$ via the relationship $c_{s}=\sqrt{\hbar g \left|\psi\right|^2/m_{pol}}$ (22) , where $g$ is the polariton-polariton interaction constant.

Figure~\ref{Fig1}A shows the image of a polariton fluid with $k=0.73~\mu m^{-1}$ and $v_{flow} = 1.7~\mu m/ps$, created with a Gaussian excitation spot of $30~\mu m$ in diameter. The resonant pump is centred slightly upstream from a photonic defect of $4.5~\mu m$ present in the microcavity, in order not to lock the phase of the flowing condensate past the defect. Two oblique dark solitons with a width of $3-5~\mu m$ are spontaneously generated in the wake of the barrier created by the defect, and propagate within the polariton fluid in a straight line (Fig.~\ref{Fig1}B).

An unambiguous characteristic of solitons in BECs is the phase jump across the soliton (12, 13, 26). In order to reveal the phase variations in the polariton quantum fluid we make the emission from the condensate interfere with a reference beam of homogeneous phase, with a given angle between the two beams (20). The result (Fig.~\ref{Fig1}C) shows a phase jump of up to $\pi$ (half an interference period) as a discontinuity in the interference maxima along the soliton.

The one-dimensional soliton relationships obtained from the solution of the Gross-Pitaevskii equation (13, 26) can be extended to two-dimensions to relate the soliton velocity $v_s$ in the reference frame of the fluid, the phase jump $\delta$, and depth $n_{s}$ with respect to the polariton density $n$ away from the soliton:

\begin{align}
cos\left( \frac{\delta}{2}\right)=\left( 1-\frac{n_{s}}{n}\right)^{1/2}=\frac{v_{s}}{c_{s}}
\label{equation1}
\end{align}

\begin{figure}[t]
	{\includegraphics[width=\columnwidth]{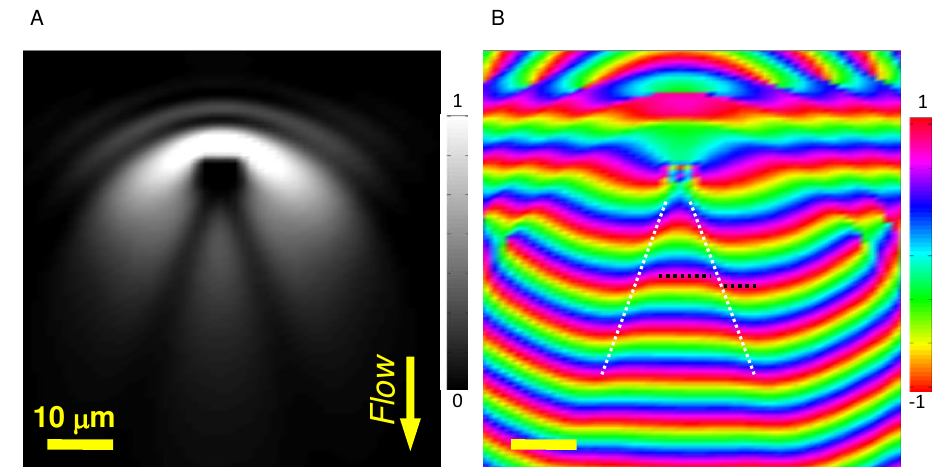}}
	\caption{
(A) Real space emission obtained from the solution of the non-equilibrium Gross-Pitaevskii equation for the parameters of the experiment depicted in Fig.~\ref{Fig1}. (B) Normalized real part of the polariton wavefunction, showing a phase jump (dark dashed lines) along the solitons (white dotted lines).
}
	\label{Fig2}
\end{figure}

In our geometry, a soliton standing in a straight line in the laboratory frame implies a constant $v_{s}=v_{flow} \sin(\alpha)$, where $\alpha$ is defined in Fig.~\ref{Fig1}A. As the soliton becomes darker ($n_{s}$ approaching $n$) the phase jump saturates at $\delta=\pi$. Indeed, the solitons remain quite deep up to the first $40~\mu m$ of trajectory (Fig.~\ref{Fig1}D), with a corresponding phase jump close to, but smaller than $\pi$. At longer distances, the depth decreases along with the phase jump. Open triangles in Fig.~\ref{Fig1}D show the ratio $n_{s}/n$ as obtained from the measured phase jump and Eq.~1. This confirms that the soliton relationships, which were derived for condensates without dissipation (26) are applicable locally to the case of polaritons under cw pumping, where the polariton density is stationary in time. Note that the polariton density continuously decreases downstream from the barrier due to the finite polariton lifetime. This results in a decrease of the speed of sound (from $c_{s}=3.5\pm 1~\mu m/ps$ at $\Delta y=14~\mu m$, to $c_{s}=1.2\pm 0.5~\mu m/ps$ at $\Delta y=50~\mu m$, see (25) for the estimation of $c_{s}$), which compensates the expected acceleration of the soliton when it becomes less deep (smaller $n_{s}/n$ in Eq.~1). As a consequence, the solitons present an almost rectilinear shape.

Simulations based on the Gross-Pitaevskii equation with pumping and decay (25) according to the model described in (24) for the experimental parameters of Fig.~\ref{Fig1} show the nucleation of a pair of solitons (Fig.~\ref{Fig2}A) with its associated phase jump (Fig.~\ref{Fig2}B). The model confirms that dark solitons nucleate hydrodynamically due to the gradient of flow speeds occurring around the potential barrier, which result in density variations on the order of the healing length. Once the soliton is formed, the repulsive interparticle interactions stabilize its shape as it propagates (6, 27-29). In contrast, no stable soliton was observed at low excitation density when polariton-polariton interactions are negligible (see Fig. 7).

\begin{figure*}[t]
	{\includegraphics[width=0.7 \textwidth]{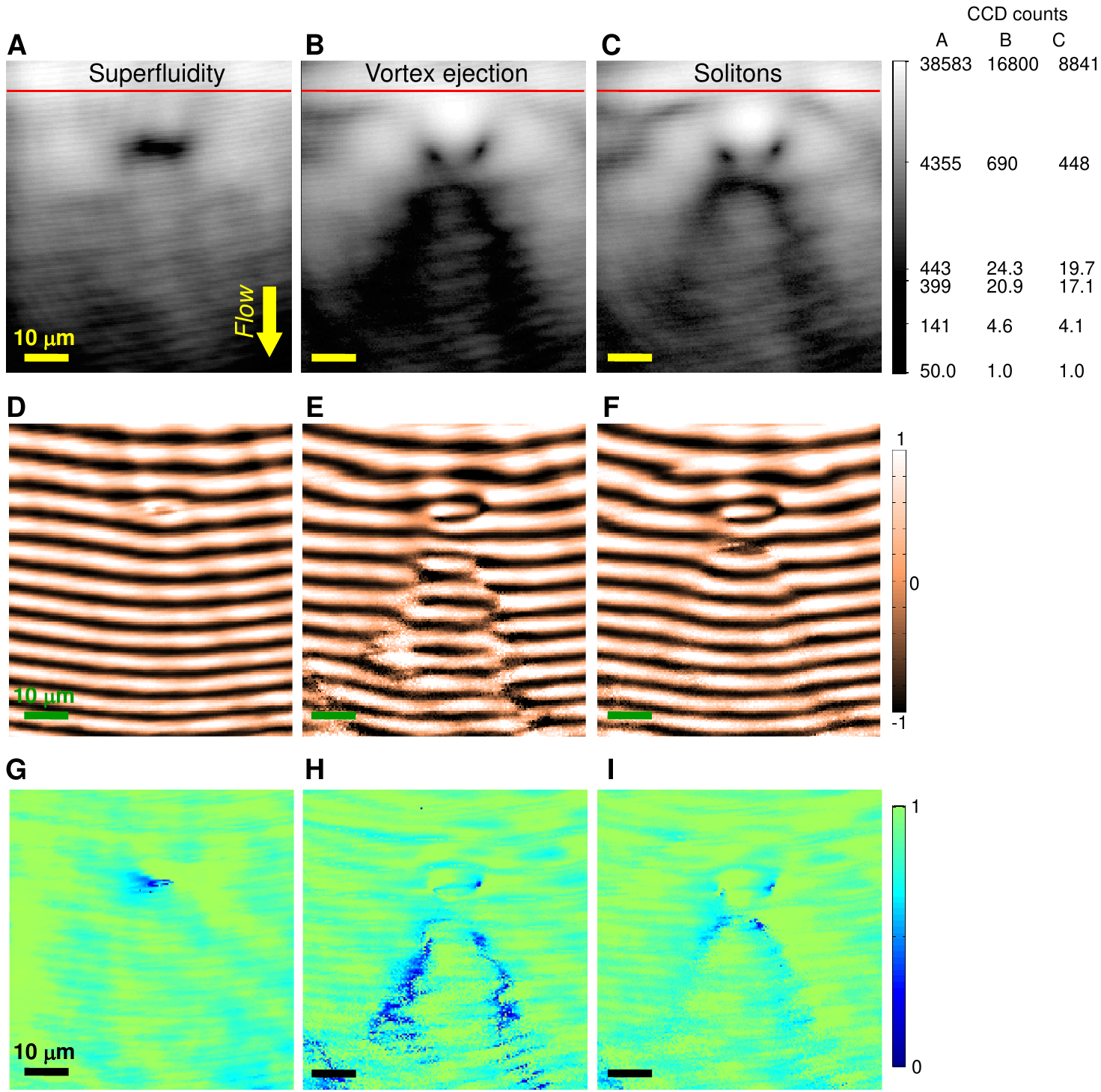}}
	\caption{
(A)-(C) Real space images of the polariton gas flowing downward at different excitation densities in the presence of a double defect (total width: $15~\mu m$). The gas is injected above the red line (25). At high density (A) (117 mW), the fluid is subsonic ($v_{flow}=0.25\bar{c}_{s}$) and flows in a superfluid fashion around the defect. At lower densities (B) (36 mW; $v_{flow}=0.4\bar{c}_{s}$), a turbulent pattern appears in the wake of the defect eventually giving rise to the formation of two oblique dark solitons (C) (27 mW; $v_{flow}=0.6\bar{c}_{s}$). (D)-(F) Interferograms corresponding to (A)-(C), respectively. (G)-(I) Show the corresponding degree of first order coherence ($g^{(1)}$, see (25)).
}
	\label{Fig3}
\end{figure*}

Other hydrodynamic regimes can be explored by varying the mean polariton density (i.e., the speed of sound) for a fixed flow speed, as shown in Fig.~\ref{Fig3}. Here, polaritons move slower than in Fig.~\ref{Fig1} ($v_{flow}=0.79~\mu m/ps$, $k=0.34~\mu m^{-1}$), and due to their limited lifetime they cannot propagate far away from the excitation spot. For this reason, we  have designed an excitation spot with the shape of half a Gaussian, with an abrupt intensity cut off (see Fig. 5). Below the red line in Fig.~\ref{Fig3}A-C, only polaritons propagating away from the pumped area are present, and their phase is not imposed by the resonant pump beam.

Figure~\ref{Fig3}A shows the polariton flow at subsonic speeds ($v_{flow}=0.25\bar{c}_{s}$, where the bar indicates the mean speed of sound), at high excitation density. The condensate is in the superfluid regime as evidenced from the absence of density modulations in the fluid hitting the barrier and from the homogeneous phase (Fig.~\ref{Fig3}D), showing a high value of  the zero time first order coherence (25), $g^{(1)}$ (Fig.~\ref{Fig3}G). When the excitation density and, correspondingly, the sound speed is decreased to $v_{flow}=0.4\bar{c}_{s}$ (Fig.~\ref{Fig3}B), the fluid enters into a regime of turbulence characterised by the appearance of two low density channels in the wake created by the barrier, with extended phase dislocations (Fig.~\ref{Fig3}E). We interpret this regime as corresponding to the continuous emission of pairs of quantised vortices and antivortices moving through those channels (4-6, 24). Although a direct observation of the phase singularity of the emitted vortices is not possible under time integrated CW experiments, the effects of the vortex flow are clearly seen when looking at $g^{(1)}$. Figure~\ref{Fig3}H shows a trace of low degree of coherence along each channel, due to the continuous passage of individual vortices. Finally, if the density is further decreased, we observe the formation of oblique dark solitons (Fig.~\ref{Fig3}C; $v_{flow}=0.6\bar{c}_{s}$), with the characteristic phase jump along their trajectory (Fig.~\ref{Fig3}F), and a constant value of $g^{(1)}$ close to 1 (Fig.~\ref{Fig3}I). 

The three regimes depicted in Fig.~\ref{Fig3} have been anticipated by the non-equilibrium Gross-Pitaevskii model (24). We report a break up of the superfluid regime at $v_{flow} \sim 0.4\bar{c}_{s}$, a value consistent with predictions for the onset of drag in the presence of large circular barriers (4, 5). Our observations show that solitons in the polariton fluid can be stable down to subsonic speeds. This is in contrast to calculations for atomic condensates, in which oblique dark solitons are predicted to be stable only at supersonic speeds (6, 27). Since our non-equilibrium simulations (Fig.~\ref{Fig2}) reproduce the observed nucleation at subsonic speeds, we infer that the additional damping in the polariton system arising from the finite lifetime is responsible for the stabilization of the soliton at subsonic speeds.

\begin{figure}[t]
	{\includegraphics[width=\columnwidth]{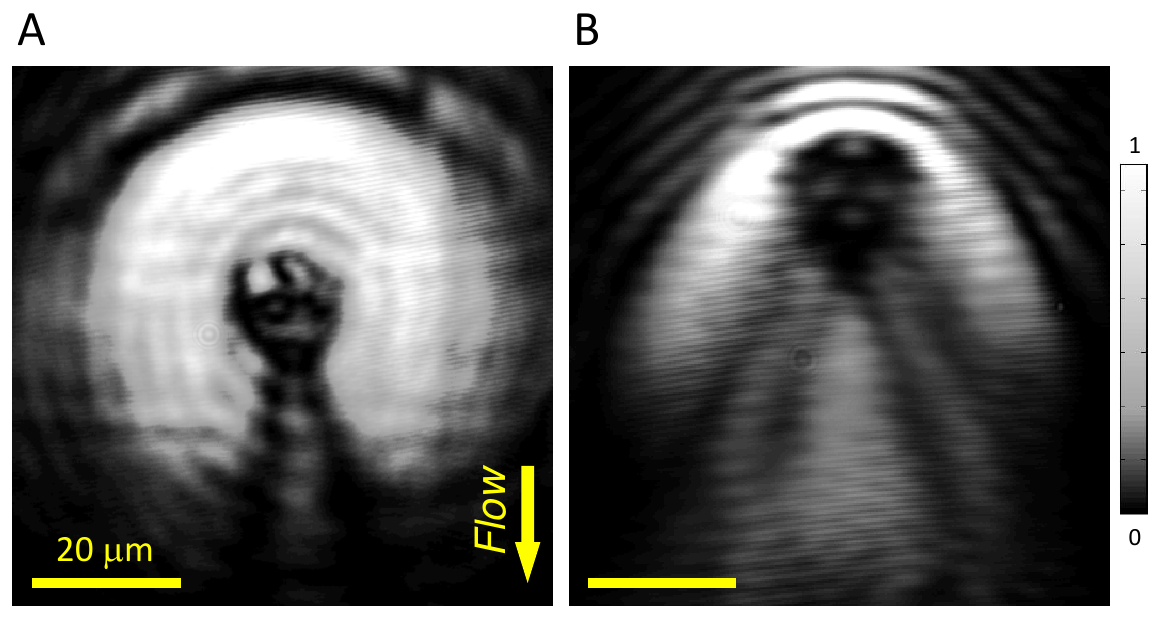}}
	\caption{Real space images of the polariton flow around a large defect ($17~\mu m$ in diameter) at low (A) ($k=0.2~\mu m^{-1}$) and high (B) ($k=1.1~\mu m^{-1}$) injected wavevectors showing, respectively, the formation of a soliton doublet and quadruplet.
}
	\label{Fig4}
\end{figure}

Finally, we have explored the possibility of going beyond the generation of soliton doublets by using a large circular potential barrier (6). Figure~\ref{Fig4}A shows a polariton flow at low momentum ($k=0.2~\mu m^{-1}$) injected in a Gaussian spot slightly above the obstacle, which nucleates a soliton doublet. If the momentum of the flow is increased above a certain value, the strong density mismatch before and after the defect is able to generate a soliton quadruplet (Fig.~\ref{Fig4}B, $k=1.1~\mu m^{-1}$). In principle, it should be possible to access even higher order solitons by increasing both the obstacle size and the ratio $v_{flow}/c_{s}$.

Our results demonstrate the potential of polariton superfluids for experimental studies of quantum hydrodynamics. Both the velocity and the density of the quantum fluid can be finely controlled by optical means, and simultaneous access to the condensate density, phase and coherence is available from the emitted light. These features have been essential in the reported observation of hydrodynamic generation of oblique solitons in the wake of potential barriers, and offer the opportunity to probe more complex phenomena like Andreev reflections (30), nucleation and trapping of vortex lattices (24), and quantum turbulence (31).

We thank S. Barbay, J. Bloch, R. Kuszelewicz, W.D. Phillips, L.P. Pitaevskii, and M. Wouters for useful discussions, and L. Martiradonna for the confocal masks. This work was supported by the IFRAF, CLERMONT4 and the Agence Nationale pour la Recherche. A.B. and C.C. are members of the Institut Universitaire de France.

Raw data from all figures can be accessed as ancillary files.

\subsection*{References and Notes}
\begin{itemize}
\item[1.]
A. J. Leggett, \textit{Rev. Mod. Phys.} \textbf{71}, S318 (1999).

\item[2.]
I. Carusotto, S. X. Hu, L. A. Collins, A. Smerzi, \textit{Phys. Rev. Lett.} \textbf{97}, 260403 (2006).

\item[3.]
A. Amo et al., \textit{Nature Phys.} \textbf{5}, 805 (2009).

\item[4.]
T. Frisch, Y. Pomeau, S. Rica, \textit{Phys. Rev. Lett.} \textbf{69}, 1644 (1992).

\item[5.]
T. Winiecki, B. Jackson, J. F. McCann, C. S. Adams, \textit{J. Phys. B: At. Mol. Opt. Phys.} \textbf{33}, 4069 (2000).

\item[6.]
G. A. El, A. Gammal, A. M. Kamchatnov, \textit{Phys. Rev. Lett.} \textbf{97}, 180405 (2006).

\item[7.]
C. Raman et al., \textit{Phys. Rev. Lett.} \textbf{83}, 2502 (1999).

\item[8.]
R. Onofrio et al., \textit{Phys. Rev. Lett.} \textbf{85}, 2228 (2000).

\item[9.]
P. Engels, C. Atherton, \textit{Phys. Rev. Lett.} \textit{99}, 160405 (2007).

\item[10.]
S. Inouye et al., \textit{Phys. Rev. Lett.} \textbf{87}, 080402 (2001).

\item[11.]
T. W. Neely, E. C. Samson, A. S. Bradley, M. J. Davis, B. P. Anderson, \textit{Phys. Rev. Lett.} \textbf{104}, 160401 (2010).

\item[12.]
S. Burger et al., \textit{Phys. Rev. Lett.} \textbf{83}, 5198 (1999).

\item[13.]
J. Denschlag et al., \textit{Science} \textbf{287}, 97 (2000).

\item[14.]
J. J. Chang, P. Engels, M. A. Hoefer, \textit{Phys. Rev. Lett.} \textbf{101}, 170404 (2008).

\item[15.]
J. Kasprzak et al., \textit{Nature} \textbf{443}, 409 (2006).

\item[16.]
S. Christopoulos et al., \textit{Phys. Rev. Lett.} \textbf{98}, 126405 (2007).

\item[17.]
A. Amo et al., \textit{Nature} \textbf{457}, 291 (2009).

\item[18.]
E. Wertz et al., \textit{Nature Phys.} \textbf{6}, 860 (2010).

\item[19.]
A. Amo et al., \textit{Phys. Rev. B} \textbf{82}, 081301 (2010).

\item[20.]
K. G. Lagoudakis et al., \textit{Nature Phys.} \textbf{4}, 706 (2008).

\item[21.]
C. W. Lai et al., \textit{Nature} \textbf{450}, 529 (2007).

\item[22.]
I. Carusotto, C. Ciuti, \textit{Phys. Rev. Lett.} \textbf{93}, 166401 (2004).

\item[23.]
D. Sanvitto et al., \textit{Nature Phys.} \textbf{6}, 527 (2010).

\item[24.]
S. Pigeon, I. Carusotto, C. Ciuti, \textit{Phys. Rev. B} \textbf{83}, 144513 (2011).

\item[25.]
See materials and methods at the end of the manuscript.

\item[26.]
A. D. Jackson, G. M. Kavoulakis, C. J. Pethick, \textit{Phys. Rev. A} \textbf{58}, 2417 (1998).

\item[27.]
A. M. Kamchatnov, L. P. Pitaevskii, \textit{Phys. Rev. Lett.} \textbf{100}, 160402 (2008).

\item[28.]
A. V. Yulin, O. A. Egorov, F. Lederer, D. V. Skryabin, \textit{Phys. Rev. A} \textbf{78}, 061801 (2008).

\item[29.]
Y. Larionova, W. Stolz, C. O. Weiss, \textit{Opt. Lett.} \textbf{33}, 321 (2008).

\item[30.]
A. J. Daley, P. Zoller, B. Trauzettel, \textit{Phys. Rev. Lett.} \textbf{100}, 110404 (2008).

\item[31.]
N. G. Berloff, preprint avalilable at \textit{arXiv:1010.5225 (2010)}.

\end{itemize} 

%\begin{thebibliography}
%%\bibitem{Anderson95}
%M. H. Anderson et al., Science \textbf{269} 198 (1995)
%
%%\bibitem{Davis95}
%K. B. Davis et al., Phys. Rev. Lett. \textbf{75} 3969 (1995)
%\end{thebibliography}
%\bibliography{library}

\newpage

\section{Materials and methods}
\subsection{Sample description}
Our sample is a $3\lambda/2$ GaAs cavity with three In$_{0.05}$Ga$_{0.95}$As quantum wells resulting in a Rabi splitting of $5.1~meV$, and a polariton lifetime of about $15~ps$. The top/bottom distributed Bragg reflectors forming the cavity have 21/24 pairs of GaAs/AlGaAs alternating layers with an optical thickness of $\lambda/4$, $\lambda$ being the wavelength of the energy of the confined cavity mode. All our experiments are performed at zero exciton-cavity detuning, with a continuous wave single mode laser quasi-resonant with the lower polariton branch.

The sample has been grown by molecular beam epitaxy. During the growth of the distributed Bragg reflectors, the slight lattice mismatch between the materials of each layer results in an accumulated stress which relaxes in the form of structural defects. These photonic defects form a very high potential barrier in the polariton energy landscape.

\subsection{Confocal excitation scheme}
The data reported in Fig. 3 have been taken making use of the confocal excitation scheme represented in Fig. 5. The laser is focalised in an intermediate plane where a mask is placed in order to hide the upper part of the Gaussian spot on that plane. Then, an image of the intermediate plane is done on the sample, producing a spot with the shape of a half Gaussian (the profile is depicted in the inset of Fig. 5). Polaritons are resonantly injected in the microcavity with a well-defined wavevector, in the region above the red line in the figure. In these conditions, polaritons move out of the excitation spot with a free phase, not imposed by the pump beam. This is essential for the observation of hydrodynamic effects involving topological excitations with phase discontinuities.

\begin{figure}[t]
	{\includegraphics[width=\columnwidth]{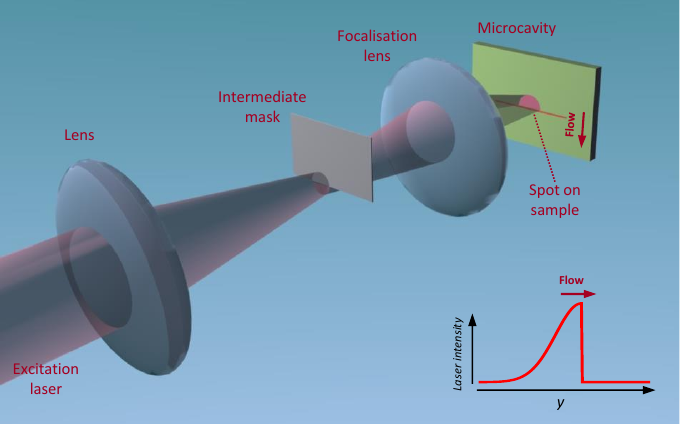}}
	\caption{Excitation setup used for the experiments reported in Fig. 3. The intermediate mask creates a spot on the sample with the shape of a half Gaussian (the inset shows a $y$ cross-section of the spot).
}
	\label{FigS1}
\end{figure}

\subsection{Estimation of the sound speed}
The average sound speeds reported in the main manuscript have been obtained from the measured soliton speed $v_{s}$ and phase jump $\delta$, with the use of Eq. 1 ($\cos(\delta/2)=v_{s}
/c_{s}$). In Figs. 1 and 3 we have estimated the sound speed in the soliton regime (Fig. 1a, 1c and Fig. 3c, 3f) in the region below the potential barrier, where the hydrodynamic effects are observed. We have taken as the soliton speed $v_{s}=v_{flow}\sin(\alpha)$ , where $2\alpha$ is the angle of aperture of the soliton pair, and $v_{flow}$ is obtained from the injected polariton wavevector and the measured polariton mass via $v_{flow}=k\hbar /m_{pol}$.

In the case of Fig. 3, the sound speed is estimated from the phase jump at half the total propagation distance in the soliton regime (Fig. 3c, 3f). In order to obtain the sound speed for other two excitation densities (panels a,b,d,e,g,h), we use the measured polariton density relative to the soliton case (c,f) and the sound speed relation $c_{s}=\sqrt{\hbar g \left|\psi\right|^2/m_{pol}}$. Note that the sound speed is proportional to the square root of the density $\left|\psi\right|^2$.

In order to confirm that this relationship is consistent with our results, we proceed in the same way for the data plotted in Fig.~1. In this case we take the sound speed obtained from the phase jump along the right soliton. The sound speed decays as the fluid is further away from the excitation area. The result is shown in black dots in Fig.~6. Additionally, we measure the decay of the density on the edges of the soliton along the soliton line. In Fig.~6 we plot in green triangles the magnitude $c_{s}=A \sqrt(I)$ , where $I$ is the emitted intensity (proportional to the polariton density) and $A$ is a fitting constant. The figure shows that the decay of the sound speed obtained from both the phase jump and the measured density follow the same trend.

These results justify our method to obtain the sound speed in the superfluid and vortex emission regimes (Fig. 3a,b) from the measured sound speed in the soliton regime (Fig. 3c, obtained from the phase jump) and the relative polariton density.

\begin{figure}[t]
	{\includegraphics[width=\columnwidth]{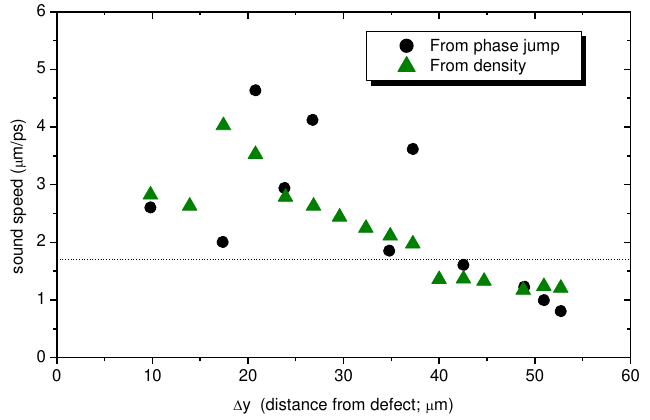}}
	\caption{Sound speed estimation for the data of Fig. 1. Black dots show the sound speed obtained from the phase jump and Eq. 1 along the soliton trajectory. Green triangles show the fit from the measured square root of the emitted intensity (proportional to the polariton density). The dashed line shows the fluid speed.).
}
	\label{FigS2}
\end{figure}

\subsection{Degree of first order coherence}

The degree of first order coherence, $g^{(1)}$, is defined as:

\begin{align}
g^{(1)} ( \mathbf{r_{1}}, \mathbf{r_{2}}, t, \tau)= \frac{\left\langle \psi^{\dagger}\left( \mathbf{r_{1}}, t \right) \psi\left( \mathbf{r_{2}}, t+\tau \right)\right\rangle}{\sqrt{\left\langle \left|\psi\left( \mathbf{r_{1}}, t \right)\right|^{2}\right\rangle \left\langle \left|\psi\left( \mathbf{r_{2}}, t+\tau \right)\right|^{2}\right\rangle}}
\label{equation2}
\end{align}

In our cw experiments in stationary conditions, $g^{(1)}$ is independent of $t$. In order to measure $g_{(1)}(\tau=0)$, we direct the emitted light from the polariton condensate, which contains all the coherence information of the wavefunction, into a modified Mach-Zehnder interferometer. The interference image is obtained from the composition of the real space emitted field with coordinate $\mathbf{r_{1}}$, and a reference beam issued from the enlarging of a small area of the emission with a fixed position $\mathbf{r_{2}}$ with a well defined spatial phase. By varying the length of the reference beam arm by up to two wavelengths around zero delay, we measure the visibility of the fringes of the interferometric image, giving direct access to the time averaged real space degree of coherence of the condensate wavefunction $\psi\left( \mathbf{r_{1}}, t \right)$ with respect to a coherent reference $\psi\left( \mathbf{r_{2}}, t \right)$.

\subsection{Gross-Piteavskii equation}

Figure 2 shows simulations based on the solution of a generalized non-equilibrium Gross-Pitaevskii equation describing the polariton condensate subject to interparticle interactions. In the basis of the confined exciton and photon wavefunctions it has the form:

	{\includegraphics[width=\columnwidth]{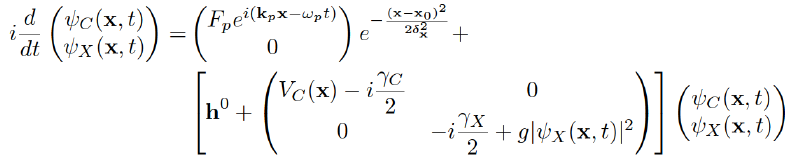}}

where, 

	{\includegraphics[width=0.4\columnwidth]{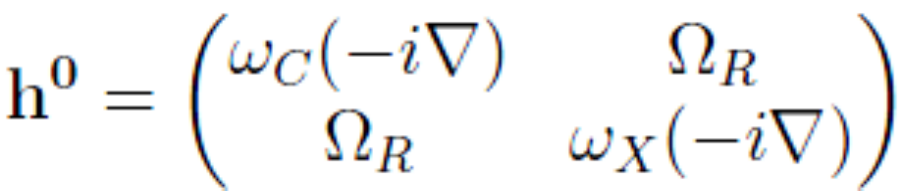}}

where, $\mathbf{x}$ is a two-dimensional spatial vector, $\psi_{X(C)}$  is the exciton (cavity photon) wavefunction, $F_{p}$, $k_{p}$ and $\hbar \omega_{p}$ are, respectively, the amplitude, momentum and energy of the pump field. The k-dependent energy of the excitons (cavity photons) is described by $\hbar \omega_{X(C)}$, $\gamma_{X(C)}$ is the decay rate of the excitons (cavity photons), with a value of 16 ps, $2\hbar \Omega_{R}$ is the vacuum Rabi splitting between the polariton modes (5.1~meV), $V_{C}\left( \mathbf{x}\right)$ is the photonic potential barrier, $g$ the exciton-exciton interaction constant, taken to be $0.01~meV~\mu m^{2}$. $\mathbf{x_0}$ indicates the position of the centre of the Gaussian spot on the sample, while $\delta_X$ is its radial width. In the simulations shown in Fig.~2, $k_{p}=0.73~\mu m^{-1}$ and the pump energy is detuned from the lower polariton branch at that $k_p$ by $0.2~meV$. The defect is simulated as a rectangle of $5\times 3~\mu m$ and a height of $80~meV$.

\newpage

\begin{figure}[t]
	{\includegraphics[width=\columnwidth]{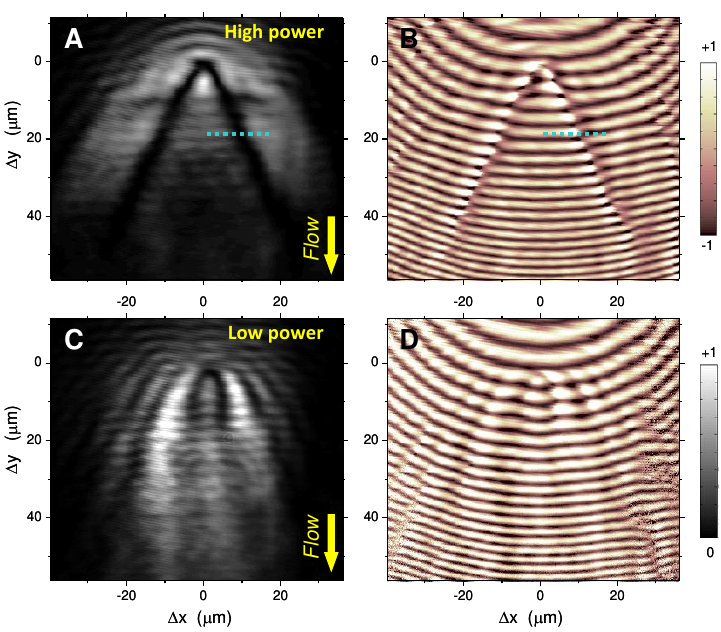}}
	\caption{(A) Real space emission showing oblique dark solitons at high excitation density ($85~mW$), and (B) the corresponding interference pattern showing the phase slip along the soliton trajectory (reproduced from Fig.~1A and C). (C and D) Real space emission and corresponding interference pattern at low excitation density ($10~mW$). In this case, polariton-polariton interactions are negligible and solitons are neither formed nor sustained in the fluid. The parabolic wave patterns observed in (C) arise from the interference between the injected polaritons and those elastically scattered by the defect. The interference pattern (D) does not show phase jumps as those associated to solitons.
}
	\label{FigS4}
	\end{figure}

%\newcolumn
\begin{figure}[t]
	{\includegraphics[width=\columnwidth]{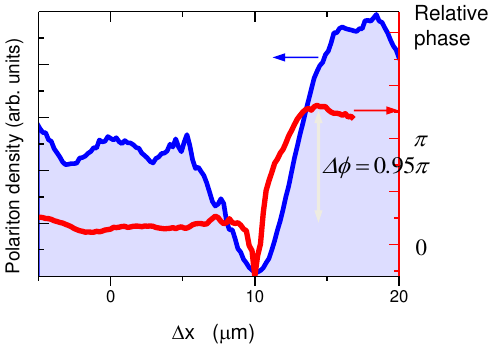}}
	\caption{Intensity (blue line) and phase profile (red line), along the dashed line indicated in Fig. 7, showing a phase jump of the condensate wavefuntion of almost $\pi$ across the the soliton.
}
	\label{FigS4}
\end{figure}

\end{document}